\documentclass[amsmath,amssymb,pre]{revtex4}
\usepackage{graphicx}

\begin{document}
\begin{large}

\title{Random sequential adsorption on Euclidean, fractal and random lattices}

\author{P. M. Pasinetti}
\affiliation{Departamento de F\'{\i}sica, Instituto de F\'{\i}sica
Aplicada, Universidad Nacional de San Luis-CONICET, Ej\'ercito de Los Andes 950, D5700HHW, San Luis, Argentina}
\author{L. S. Ramirez}
\affiliation{Departamento de F\'{\i}sica, Instituto de F\'{\i}sica
Aplicada, Universidad Nacional de San Luis-CONICET, Ej\'ercito de Los Andes 950, D5700HHW, San Luis, Argentina}
\author{P. M. Centres}
\affiliation{Departamento de F\'{\i}sica, Instituto de F\'{\i}sica
Aplicada, Universidad Nacional de San Luis-CONICET, Ej\'ercito de Los Andes 950, D5700HHW, San Luis, Argentina}
\author{G. A. Cwilich}
\affiliation{Department of Physics, Yeshiva University, 500 West 185th Street, New York, New York 10033, USA}
\author{A. J. Ramirez-Pastor}
\affiliation{Departamento de F\'{\i}sica, Instituto de F\'{\i}sica
Aplicada, Universidad Nacional de San Luis-CONICET, Ej\'ercito de Los Andes 950, D5700HHW, San Luis, Argentina}

\date{\today}

\begin{abstract}

Irreversible adsorption of objects of different shapes and sizes on Euclidean, fractal and random lattices is studied. The adsorption process is modeled by using random sequential adsorption (RSA) algorithm. Objects are adsorbed on one-, two-, and three-dimensional Euclidean lattices, on Sierpinski carpets having dimension $d$ between 1 and 2, and on Erdos-Renyi random graphs. The number of sites is $M=L^d$ for Euclidean and fractal lattices, where $L$ is a characteristic length of the system. In the case of random graphs it does not exist such characteristic length, and the substrate can be characterized by a fixed set of $M$ vertices (sites) and an average connectivity (or degree) $g$. The paper concentrates on measuring (1) the probability $W_{L(M)}(\theta)$ that a lattice composed of $L^d(M)$ elements reaches a coverage $\theta$, and (2) the exponent $\nu_j$ characterizing the so-called ``jamming transition". The results obtained for Euclidean, fractal and random lattices indicate that the main quantities derived from the jamming probability $W_{L(M)}(\theta)$ behave asymptotically as $M^{1/2}$. In the case of Euclidean and fractal lattices, where $L$ and $d$ can be defined, the asymptotic behavior can be written as $M^{1/2} = L^{d/2}=L^{1/\nu_j}$, and $\nu_j=2/d$.
\end{abstract}


\maketitle


\newpage

 \section{Introduction} \label{intro}

The deposition (or adsorption) of particles on solid surfaces has been a central research topic of statistical mechanics. In many experiments on adhesion of colloidal particles and proteins on solids substrates the relaxation time scales are much longer than the times of the formation of the deposit. This situation has encouraged the scientific community to explore research on irreversible adsorption.

A well-known example of an irreversible monolayer deposition process is the random sequential adsorption (RSA). This process has been extensively investigated in the literature \cite{Feder,Evans,Privman1,Privman2,Talbot,Cadilhe,Budinski}, showing a wide range of applications in biology, nanotechnology, device physics, physical chemistry, and materials science. In RSA processes objects are randomly, sequentially and irreversibly deposited onto an initially empty $d$-dimensional substrate or lattice. The quantity of interest is the fraction of total area, $\theta(t)$, covered in time, $t$, by the depositing particles or objects. The objects are not allowed to overlap and they are permanently fixed at their spatial positions. Under these conditions, each deposited particle affects the geometry of all later placements. Thus, the dominant effect in RSA is the blocking of the available substrate area and the limiting (jamming) coverage $\theta_j=\theta(t=\infty)$ is less than in close packing.

In the framework of the RSA, objects of different shapes and sizes have been studied for both lattice and continuum models: linear $k$-mers (particles occupying $k$ adjacent lattice sites) \cite{Redner,Bonnier,Vandewalle,Kondrat,Lebovka,Tara2012,Kondrat2017,Slutskii}, flexible polymers \cite{Paw1,Paw2}, T-shaped objects and crosses \cite{Adam}, squares \cite{Nakamura86,Nakamura,Brosilow1991,Privman1991,Rodgers,Shida,Vieira,Kriu,PRE19}, disks and rectangles \cite{Connelly,Vigil1,Vigil2,Dickman}, regular and star polygons \cite{Ciesla,Ciesla1,Ciesla2,Ciesla3,Shelke}, spherocylinders and ellipsoids \cite{Viot,Sherwood}, etc. In all cases, the limiting or jamming coverage depends strongly on the shape
and size of the depositing particles. In this line, Cie\'sla et al. performed extensive numerical simulations of the RSA of smoothed $k$-mers, spherocylinders, and ellipses \cite{Ciesla4,Ciesla5,Ciesla6}. The authors found that the highest packing fraction is obtained for ellipses having the long-to-short axis ratio of 1.85, which is the largest anisotropy among the investigated shapes.

RSA studies have been also extended to fractal substrates, which play an important role in numerous biological and chemical processes. For example, coral fractal-like structure helps them to catch plankton effectively \cite{Basillais}. Adsorption on fractal collectors might also be applied in environmental protection in designing effective water or air filters \cite{Khasanov}. In Ref. \cite{Ciesla7}, Cie\'sla and Barbasz studied the RSA of spheres on Sierpinski's triangle and carpet-like fractals ($1 < d < 2$), and on general Cantor set ($d < 1$). The fundamental properties of the system (RSA kinetics, maximal random coverage ratio and density autocorrelation function) were measured by numerical simulations. The obtained results showed that in general, most of known dimensional properties of adsorbed monolayers are valid for non-integer dimensions.


An important quantity to study the RSA process is the probability $W_L(\theta)$ to find a jamming phase. In Ref. \cite{PHYSA38}, a simulation scheme to determine jamming thresholds was introduced. The method relies on the definition of the probability $W_L(\theta)$ that a lattice composed of $L^d$ elements reaches a coverage $\theta$. The value of $\theta_j$  can be obtained from the crossing point of the curves of $W_L(\theta)$ for different lattice sizes.

The jamming probability can also be used to determine the exponent $\nu_j$ characterizing the correlation length of the system. As established in the literature \cite{Vandewalle}, $W_L(\theta)$ [$\mathrm{d} W_L(\theta)/\mathrm{d} \theta$] can be fitted by the error [Gaussian] function. Then, the maximum of the derivative of the jamming probability $[\mathrm{d} W_L(\theta)/\mathrm{d} \theta]_{max}$ and the width of the transition $\Delta$ are asymptotically proportional to $L^{1/\nu_j}$ (more details about these calculations are provided in Sec. \ref{model}). Following this theoretical approach, Vandewalle et al. \cite{Vandewalle} reported a value of $\nu_j=1.0 \pm 0.1$ for the RSA of needles on square lattices (independent of the aspect ratio of the needles). The same value was reported by Nakamura \cite{Nakamura86} for the case of $k \times k$ tiles on square lattices.


More recently, the exponent $\nu_j$ was measured for different systems in 1D, 2D and 3D Euclidean lattices \cite{JSTAT9,JSTAT7,PRE19,JTATsub,PREsub}. The obtained results reveal a simple dependence of $\nu_j$ with the dimensionality of the lattice. In order to deepen these findings, the main objective of the present paper is to extend previous work to fractal and random lattices. For this purpose, dimers have been randomly, sequentially and irreversibly deposited on two kinds of substrates: $(i)$ Sierpinski carpet fractals having dimension of a non-integer value ($1<d<2$); and $(ii)$ Random networks such as the Erdos-Renyi and the Random Regular random graphs, where each graph (or lattice) is characterized by a fixed set of $M$ vertices (sites) and an average connectivity $g$. In these systems, extensive numerical simulations supplemented by analysis using finite-size scaling theory have been carried out.


The present work is a natural extension of our previous research in the area of RSA model and the results obtained here could contribute to understanding the nature of the jamming phenomenon. The paper is organized as follows: the model and main results obtained in Euclidean lattices are given in Sec. \ref{model}. The RSA problem of dimers on fractal and random lattices is presented in Sec. \ref{fractal} and Sec. \ref{random}, respectively. Finally, the conclusions are drawn in Sec. \ref{conclu}.

 \section{Model and basic definitions} \label{model}

The irreversible deposition of objects larger than a simple monomer (particle occupying one lattice site) involves the possibility of jamming. Namely, due to the blocking of the lattice by the already randomly adsorbed elements, the limiting or jamming coverage, $\theta_j=\theta(t=\infty)$ is less than that corresponding to the close packing ($\theta_j<1$). Note that $\theta(t)$ represents the fraction of lattice sites covered at time $t$ by the deposited objects. Consequently, $\theta$ ranges from $0$ to $\theta_j$ for objects occupying more than one site \cite{Evans}.

As mentioned in Sec. \ref{intro}, the jamming coverage can be obtained from the intersection of the curves of jamming probability $W_L(\theta)$ for different values of $L$. In addition, $W_L(\theta)$ can be used to determine the exponent $\nu_j$ characterizing the so-called ``jamming transition" \cite{Vandewalle}. The procedure is described below.

Let us start by defining the probability $W_L(\theta)$ that a lattice composed of $L \times L$ elements (an $L$-lattice) reaches a coverage $\theta$ \cite{PHYSA38}. In the simulations, the procedure to determine $W_L(\theta)$ consists of the following steps: (a) the construction of the $L$-lattice (initially empty) and (b) the deposition of objects on the lattice up to the jamming limit $\theta_j$. In the late step, the quantity $m_i(\theta)$ is calculated as
\begin{equation}\label{mi}
m_i(\theta)=\left\{
\begin{array}{cc}
1 & {\rm for}\ \ \theta \leq \theta_j \\
0  & {\rm for}\ \ \theta > \theta_j .
\end{array}
\right.
\end{equation}
$n$ runs of such two steps (a)-(b) are carried out for obtaining the number $m(\theta)$ of them for which a lattice reaches a coverage $\theta$,
\begin{equation}\label{m}
m(\theta) = \sum_{i=1}^n m_i(\theta).
\end{equation}
Then, $W_L(\theta)= m(\theta)/n$  is defined and the procedure is repeated for different values of $L$. Finally, it is useful now to define the quantity $W'_{L}=1-W_{L}$, which can be fitted by the error function \cite{Vandewalle},
\begin{equation}\label{error}
W'_L(\theta) = \frac{1}{\sqrt{2 \pi}\Delta_L} \int_{-\infty}^{\theta} \exp{\left[ -\frac{1}{2} \left( \frac{\theta'-\theta_j(L)}{\Delta_L} \right)^2 \right]} \mathrm{d} \theta',
\end{equation}
where $\theta_j(L)$ is the concentration at which the slope of $W'_L(\theta)$ reaches its maximum and $\Delta_L$ is the standard deviation from $\theta_j(L)$. The assumption that the distribution of critical points is a Gaussian is known not to be a Gaussian in all range of coverage \cite{Newman}. However, this quantity is approximately Gaussian near the peak, and fitting with a Gaussian function is sufficient from a practical point of view. Then, according to the finite-size scaling theory \cite{Stauffer,Vandewalle}, it is expected that,
\begin{equation}\label{derivada}
  \left(\frac{dW'_L}{d\theta}\right)_{\rm max}\propto L^{1/\nu_j},
\end{equation}
and
\begin{equation}\label{Delta}
  \Delta_{L}\propto L^{-1/\nu_j}.
\end{equation}

\begin{figure}
\begin{center}
\includegraphics[width=0.3\columnwidth]{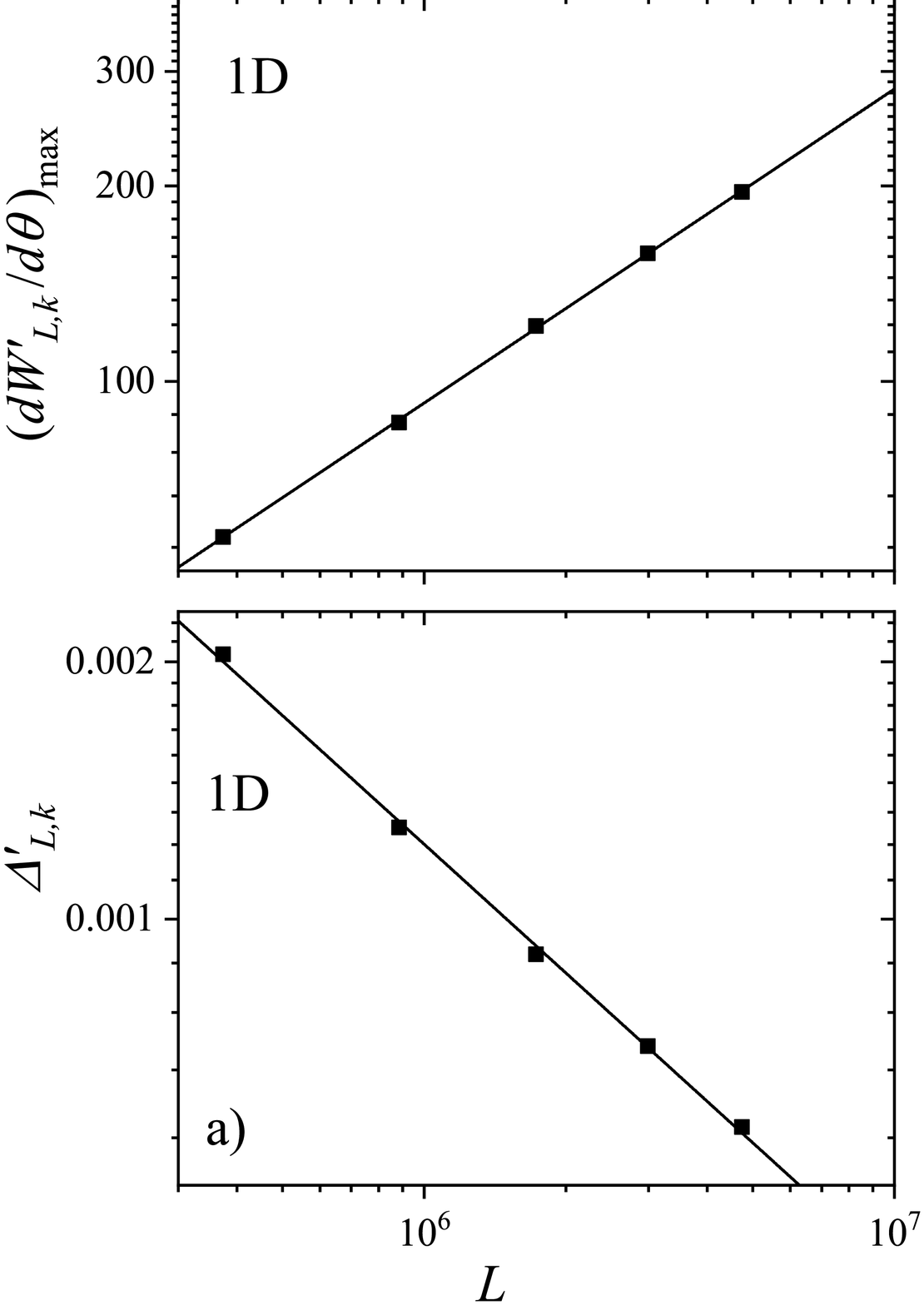}
\includegraphics[width=0.3\columnwidth]{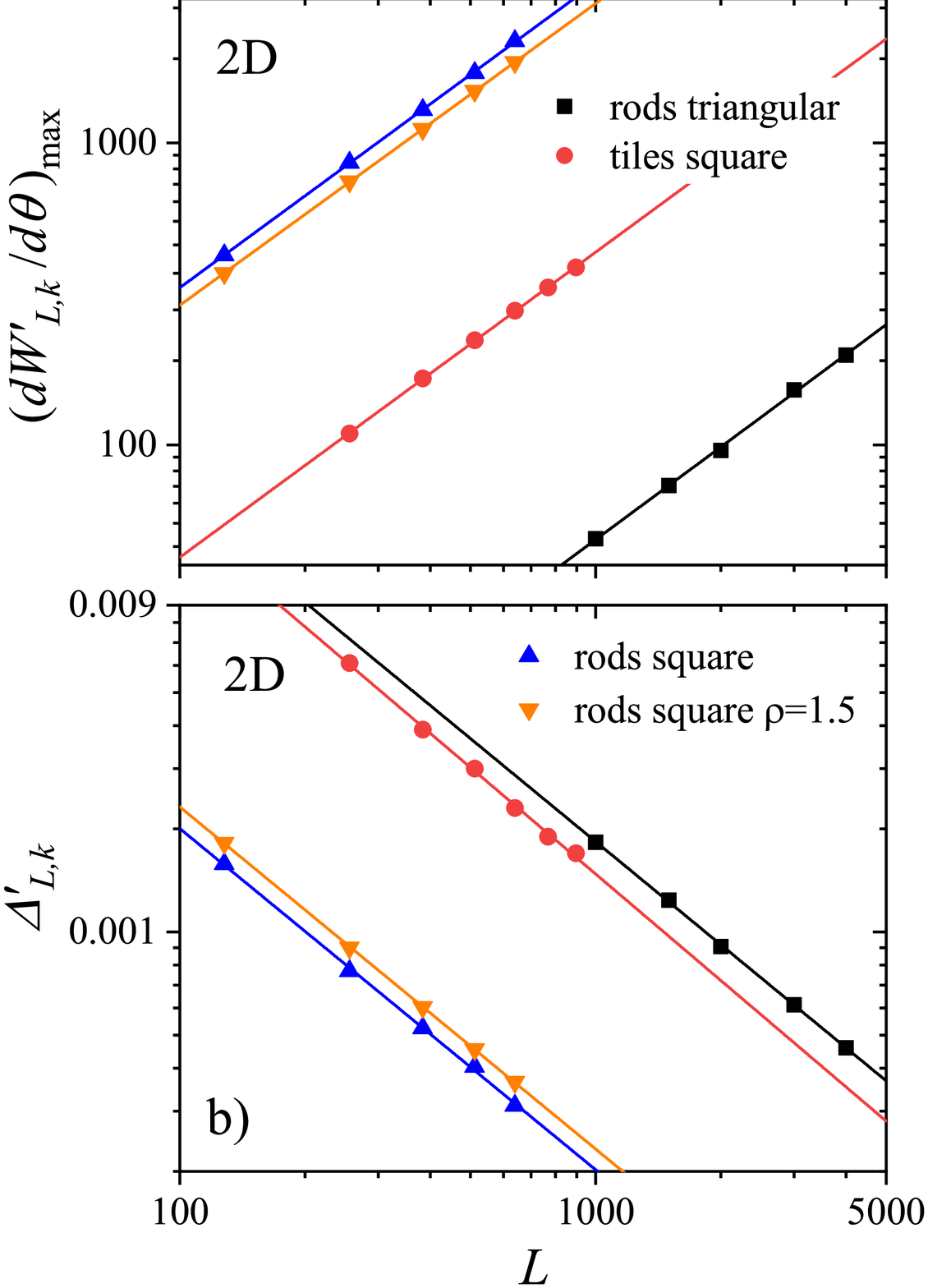}
\includegraphics[width=0.3\columnwidth]{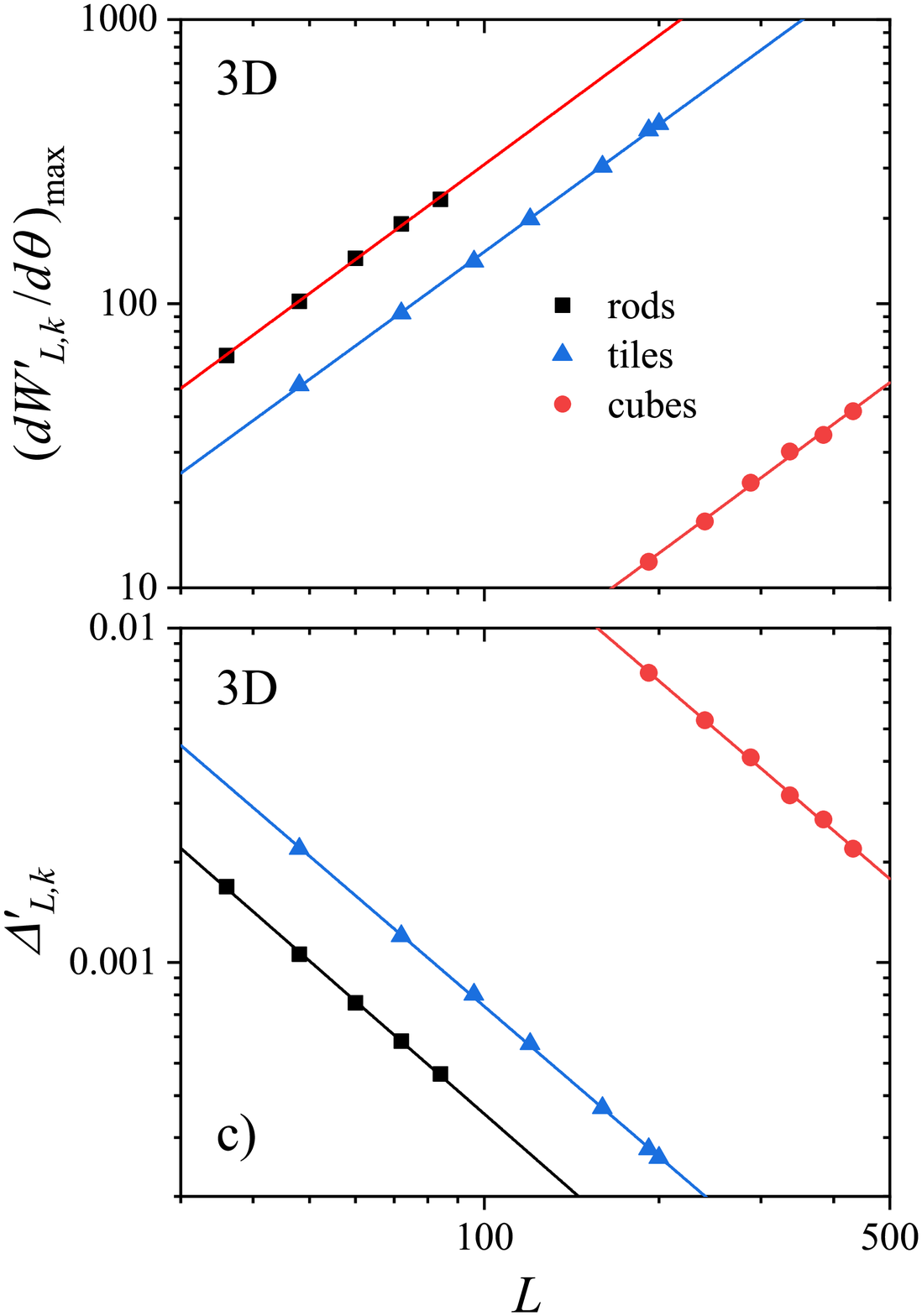}

\caption{Log-log plots of $(dW'_{L}/d\theta)_{\rm max}$ and $\Delta'_{L}$ as a function of $L$ for the following systems: a) linear $k$-mers ($k=2$) on 1D lattices; b) linear $k$-mers ($k=2$) on 2D square lattices; linear $k$-mers ($k=2$) on 2D square lattices in the presence of impurities ($\rho=0.15$); linear $k$-mers ($k=10$) on 2D triangular lattices;  $k^2$-mers ($k=4$) on 2D square lattices; c) linear $k$-mers ($k=2$) on 3D simple cubic lattices; $k^2$-mers ($k=24$) on 3D simple cubic lattices; and $k^3$-mers ($k=4$) on 3D simple cubic lattices. According to Eq. (\ref{derivada}) the slope of each line corresponds to 1/$\nu_{j}$ (or to -1/$\nu_{j}$ in the case of Eq. (\ref{Delta})).
\label{figregular}}
\end{center}
\end{figure}

Following the scheme given in Eqs. (\ref{mi}-\ref{Delta}), different systems were characterized in previous work from our group: $(1)$ linear $k$-mers on 1D lattices \cite{PHYSA38}; $(2)$ linear $k$-mers on 2D square lattices with and without the presence of impurities \cite{JSTAT3,JSTAT9}; $(3)$ linear $k$-mers on 2D triangular lattices \cite{JSTAT7}; $(4)$ $k \times k$ square tiles ($k^2$-mers) on 2D square lattices \cite{PRE19}; $(5)$ linear $k$-mers on 3D simple cubic lattices \cite{PHYSA38}; $(6)$ $k^2$-mers on 3D simple cubic lattices \cite{JTATsub} and $(7)$ $k \times k \times k$ cubic objects $k^3$-mers on 3D simple cubic lattices \cite{PREsub}. In all cases, $\nu_{j}$ was determined from Eqs. (\ref{derivada}) and (\ref{Delta}). Some typical cases are shown in Fig. \ref{figregular} and the fitting results for the slope are collected in Table I.

\begin{table}
\label{T1}
\begin{center}
\caption{Different systems characterized in previous works.}
\begin{tabular}{p{0.6cm} |p{5cm} |p{3.5cm} |p{3.5cm} }

$d$ & System & slope from $(dW'_{L}/d\theta)_{max}$ & slope from $\Delta'_{L}$   \\
\hline
1D	& rods & 0.482(20) & -0.49(1) \\
\hline
2D	& rods on square lattice & 1.009(11) & -0.999(4) \\
	& rods on square lattice $\rho=0.15$ & 0.999(3) & -0.99(2) \\
	& tiles on square lattice & 1.01(6) & -1.03(3) \\
	& rods on triangular lattice & 1.02(3) & -0.99(1) \\
\hline
3D	& rods on simple cubic lattice & 1.50(3) & -1.52(2) \\
	& tiles on simple cubic lattice & 1.51(4) & -1.49(2) \\
	& cubes on simple cubic lattice & 1.49(1) & -1.49(1) \\
\end{tabular}
\end{center}
\end{table}

The studies described in points $(1)-(7)$ revealed that $\nu_j$ shows a simple dependence on the dimensionality of the lattice $d$: $\nu_j=2/d$. Thus, $\nu_j=2$, $1$ and $2/3$ for 1D, 2D and 3D lattices, respectively. The values of $\nu_j$ do not depend on size and shape of the depositing objects. Identical results were reported by Nakamura \cite{Nakamura} and by Vandewalle et al. \cite{Vandewalle} for the case of tiles and rods on 2D square lattices.

Previous results have been obtained for RSA processes on Euclidean lattices. It is then interesting to determine whether the functionality which connects the exponent $\nu_j$ with the dimension of the  lattice is valid for fractal substrates. This point will be addressed in the next section.

 \section{Random sequential adsorption on fractal lattices} \label{fractal}

The fractals analyzed in this work belong to the 2D Sierpinski carpet family \cite{Mandelbrot}. The general procedure used to generate Sierpinski carpets is shown in Ref. \cite{Dasgupta}. 

Here we describe briefly the method based in the particular generator patterns that we have applied in the present study (see Fig. \ref{patroncarpeta}(a). A $l \times l$ square is divided into $l^2$ equal units each corresponding to a site on a square lattice. A certain number, say $m=(l-2)^2$ of these are blocked, leaving $p=l^2-(l-2)^2=4l-4$ accessible sites. 
This is the basic unit (pattern) which is repeated self-similarly in $n$ subsequent stages of generations. As an example, Fig. \ref{patroncarpeta}(b) shows a Sierpinski carpet corresponding to $l=3$ and $n=5$. The Hausdorff dimension of these fractal substrate is $d_f = \ln(p)/ \ln(l)= \ln(4l-4)/ \ln(l)$ \cite{Dasgupta}. Since $d_f$ is given basically by the number of accessible sites and by the size of the pattern, a number of different lattices can be easily generated by this method. Table II compiles the value of $d_f$ for the fractals studied here.

\begin{figure}
\begin{center}
a) \includegraphics[width=0.08\columnwidth]{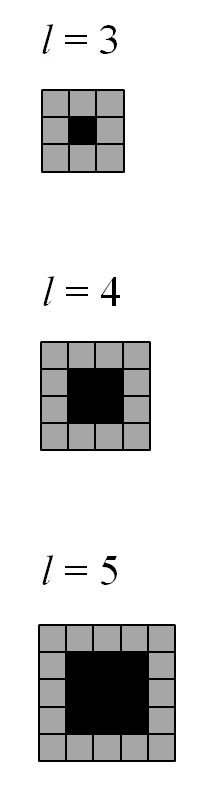} \hspace{1cm}
b) \includegraphics[width=0.3\columnwidth]{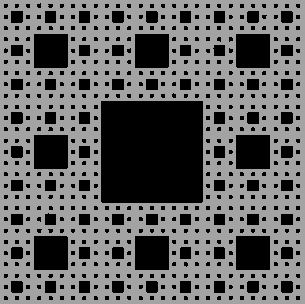} \hspace{1cm}
c) \includegraphics[width=0.3\columnwidth]{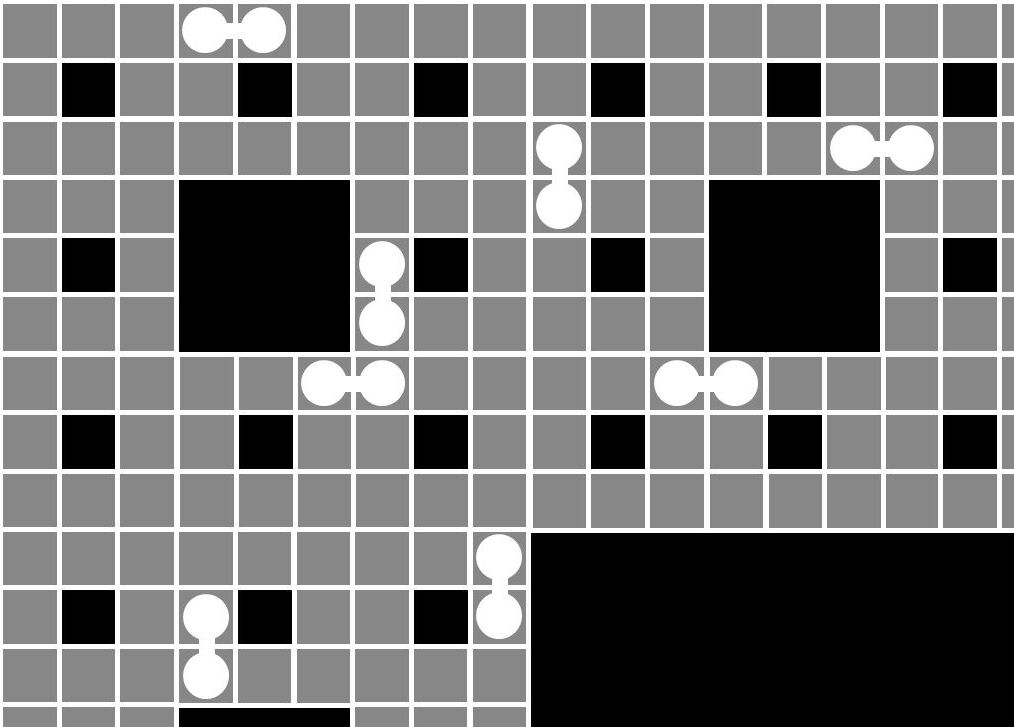} 
\caption{a) Basic units of the Sierpinski carpets studied in the present paper. b) Sierpinski carpet with $l=3$ and $n=4$. Accordingly, $L=l^n=81$ and $d_f = \ln(4l-4)/ \ln(l)=1.89279$. c) Snapshot corresponding to a typical configuration of dimers (solid circles joined by lines) on a Sierpinski carpet with $l=3$ and $n=4$. Gray squares represent accessible sites in all the figures.
\label{patroncarpeta}}
\end{center}
\end{figure}

\begin{table}
\label{T2}
\begin{center}
\caption{Hausdorff dimension ($d_f$) and linear size ($L$) for fractals generated in the present work.}
\begin{tabular}{p{1cm} |p{3.5cm} |p{3.5cm} |p{3.5cm} }
$l$ & $d_f=\ln(4l-4)/ \ln(l)$ &  slope from $(dW'_{L}/d\theta)_{max}$ & slope from $\Delta'_{L}$ \\
\hline
3	& 1.89279 & 0.928(13) & -0.937(3) \\
4	& 1.79248 & 0.908(2)  & -0.896(2) \\
5	& 1.72271 & 0.85(2)   & -0.859(2) \\
6	& 1.67195 & 0.85(1)   & -0.836(1) \\
7	& 1.63320 & 0.77(4)   & -0.812(4) \\
8	& 1.60245 & 0.802(8)  & -0.800(1) \\
\end{tabular}
\end{center}
\end{table}

After the generation of the substrate, we proceed to simulate the RSA process. In the filling process, dimers \footnote{The dimer is the simplest case of a polyatomic adsorbate and contains all the properties of the multisite-occupancy adsorption.} are deposited randomly, sequentially and irreversibly on an initially empty lattice. The deposition procedure is as follows. Given a lattice generated by patterns of size $l$, $n$ generations, linear size $L=l^n$ and $M=(4l-4)^n=L^{d_f}$  accessible sites: $(i)$ a pair of neighbouring sites is chosen at random; and $(ii)$ if the selected sites are empty and accessible, then a dimer is deposited on those sites, otherwise the attempt is rejected. A Monte Carlo step (MCs) is completed after we repeat $(i)$ and $(ii)$ $M$ times. When $N$ dimers are deposited, the concentration is  $\theta= 2 N/M$. Figure \ref{patroncarpeta}(c) shows a typical configuration corresponding to dimers (solid circles joined by lines) on a Sierpinski carpet with $l=3$ and $n = 4$.

Following the scheme described in previous section, we studied the RSA of dimers on Sierpinski carpets with different values of $l$ ($3 \leq l \leq 8 $) and $n$ ($2 \leq n \leq 5$).
The probability curves $W_{L}$ are shown in Fig. \ref{wlynuj}(a) for a typical case: $3 \times 3$ pattern and different values of $n$ between 3 and 6. Accordingly, $L=27,81,243,729$. Then, based on the jamming probability functions for various $L$, we look for $(dW'_{L}/d\theta)_{max}$ as a function of $L$ [Eq. (\ref{derivada})] and $\Delta'_{L}$ as a function of $L$ [Eq. (\ref{Delta})]. The results are shown in Fig. \ref{wlynuj}(b) in a log-log scale. In each case, $\nu_j$ can be obtained from the slope of the curve (the line in each curve is a linear fit of the points). Thus, $1/\nu_j=0.928(13)$ (maximum of the derivative) and $1/\nu_j=0.937(2)$ (standard deviation). The values obtained for $\nu_j$ coincide, within the numerical errors, with the expected value $\nu_j=2/d_f$ (in this case, $\nu_j=2/1.89279=1.05664$).

\begin{figure}
\begin{center}
\includegraphics[width=0.6\columnwidth]{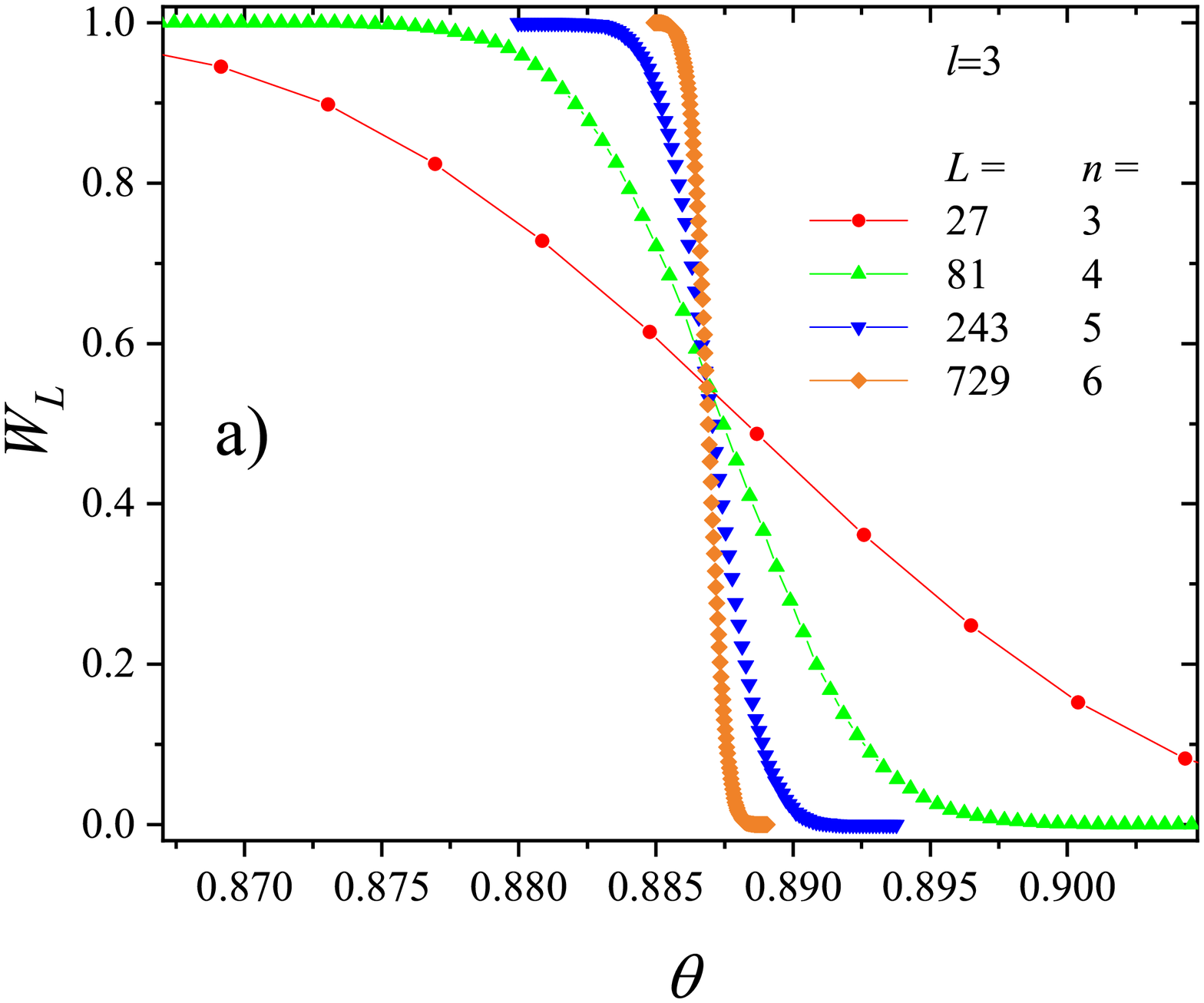}
\includegraphics[width=0.6\columnwidth]{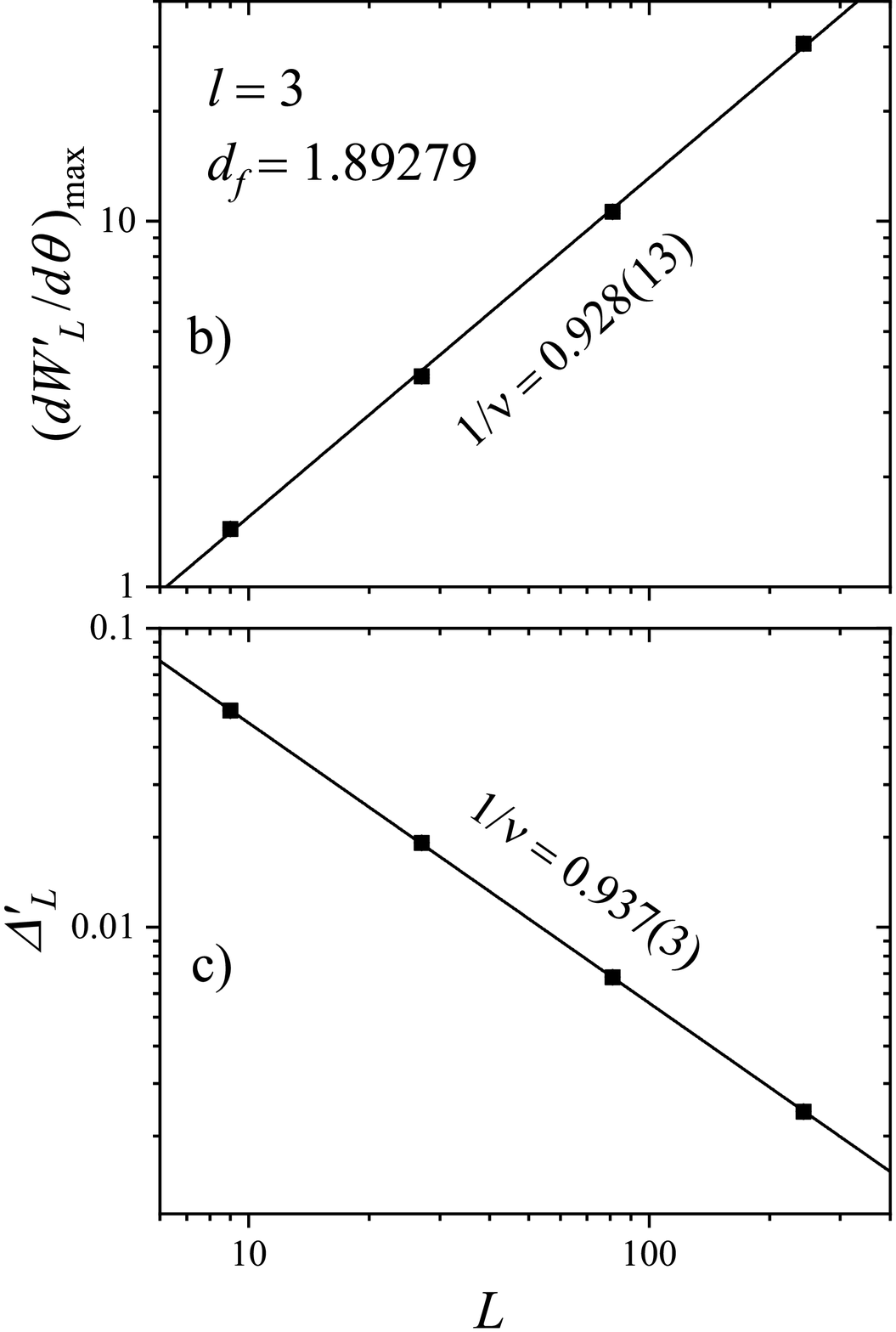}
\caption{(a) Curves of the jamming probability $W'_{L}$ as a function of the fraction of occupied sites $\theta$ for a typical case: $3 \times 3$ pattern and different values of $n$ between 2 and 5. Accordingly, the lattice sizes are $L=9,27,81,243$. (b) Log-log plot of $(dW'_{L}/d\theta)_{\rm max}$ as a function of $L$ and $\Delta'_{L}$ as a function of $L$ (inset) for the data in part (a). \label{wlynuj}}
\end{center}
\end{figure}

The procedure in Fig. \ref{wlynuj} was repeated for different values of $l=4,5,6,7,8$. In all cases, values obtained for $\nu_j$ confirm the functionality $\nu_j=2/d_f$. This situation can be clearly seen in Fig. \ref{nujfractal}, where the values of $2/\nu_j$ obtained in the present study are plotted as a function of the lattice dimension (solid circles). The previously reported values of $2/\nu_j$ for Euclidean 1D, 2D and 3D lattices are represented by open circles.

The results obtained so far allow us to rewrite Eqs. (\ref{derivada}) and (\ref{Delta}) solely in terms of the number of sites or lattice nodes, $M=L^d$. Thus,
\begin{eqnarray}\label{nodos}
  \left(\frac{dW'_L}{d\theta}\right)_{\rm max}, \ \  \Delta'_{L}   & \propto & L^{1/\nu_{j}} \nonumber \\
  & \propto & L^{d/2} = M^{1/2},
\end{eqnarray}
where, as mentioned above, $M$ is the number of sites or nodes of the lattice. In order to deepen this analysis, the next section will be devoted to the study of
RSA processes on random graphs (such as the Erdos-Renyi graphs \cite{refER1}). These substrates are characterized by a number of nodes (sites) and an (average) connectivity.

\begin{figure}
\begin{center}
\includegraphics[width=0.5\columnwidth]{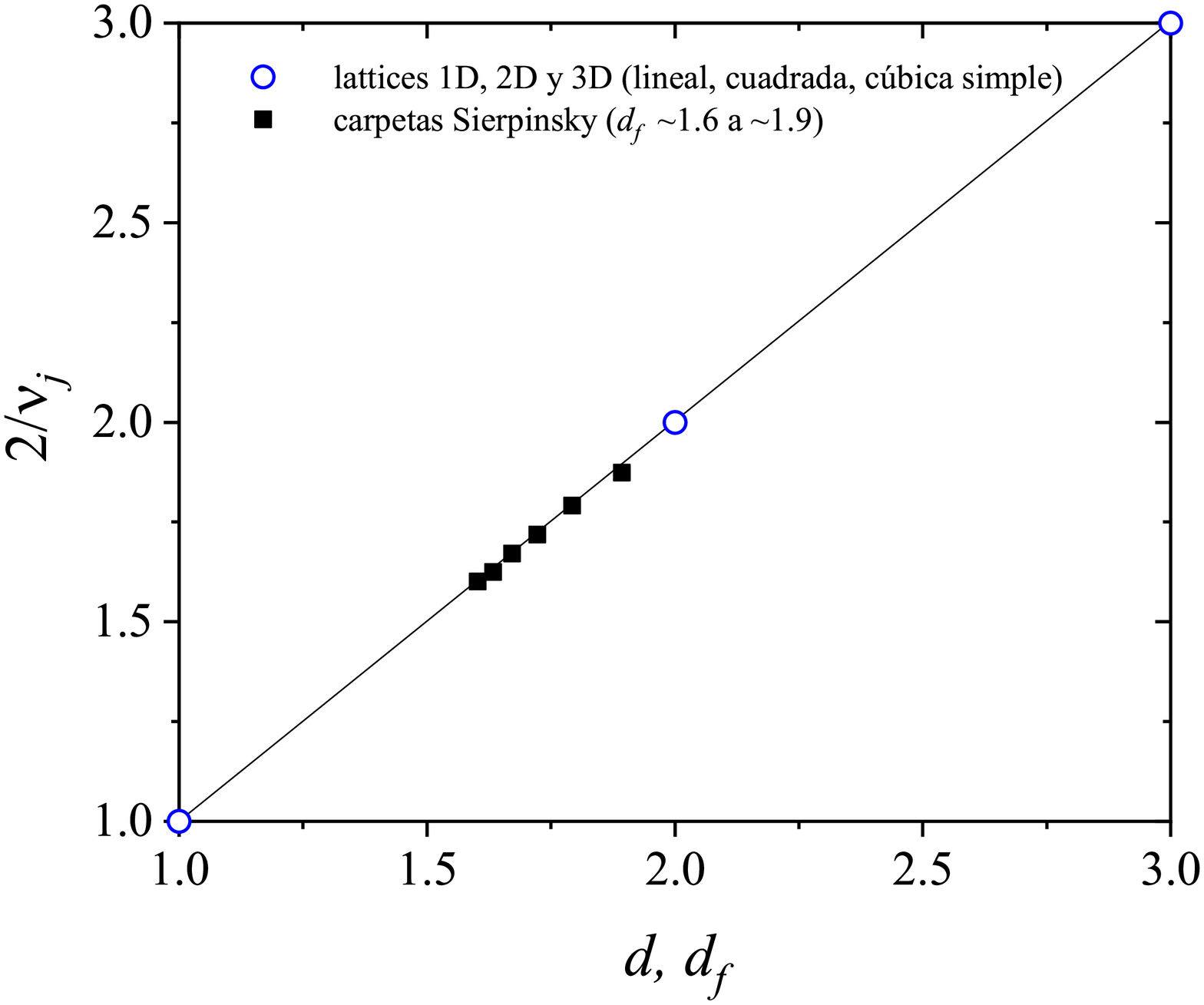}
\caption{Values of $2/\nu_j$ as a function of the lattice dimension $d$ for Euclidean 1D, 2D and 3D lattices (open circles) and Sierpinski carpets with $l$ ranging between 3 and 8 (solid squares).
\label{nujfractal}}
\end{center}
\end{figure}

 \section{Random sequential adsorption on random networks} \label{random}

The jamming process over some random networks was also studied. The random networks studied here correspond to the widely known Erdos-Renyi graph model (ER) \cite{refER1} in the form denoted as $G(M,p)$, and the Random Regular graph (RR) \cite{refRR1}. In the first case, the graph is constructed starting with $M$ initially disconnected nodes. Then, for all possible pairs of nodes, the construction process consists of connecting each pair with probability $p$, or leaving it disconnected with probability $1-p$. As it follows from the model, the nodes connectivity, or degree $g$, responds to a Poisson distribution with an average value equal to $\langle g \rangle=(M-1)p$. For the purposes of construction of the graph, $p$ is determined from the mean value of $g$, namely $p=\langle g \rangle/(M-1) \simeq \langle g \rangle/M$. In this way, the network is defined from the values of $M$ and $\langle g \rangle$. In the case of the RR graph of degree $g$, the usual construction procedure is as follows. We start with $M$ nodes with $g$ dangling (disconnected) links each. Then, the construction process consists of random selecting pairs of links and interconnecting them, until there are no more dangling links left. It is worth to note that, unlike the substrates previously considered, concepts like linear dimension, space dimension or borders, have no meaning in the case of random networks.

Once the substrates are generated, the RSA process of dimers is performed as follows.  i) A node (site) with degree $g \geq 1$ is randomly chosen. Then, one of its $g$ first neighbour sites is also chosen at random. ii) If both sites are unoccupied, a dimer is deposited occupying those sites. If not, the whole attempt is rejected. iii) We repeat the steps i) and ii) until there are no more empty spaces capable of accommodating a dimer.
The total number $N$ of dimers deposited along this procedure defines a concentration $\theta=2 N/M$ that corresponds to a particular jamming state. The whole process can be repeated a number $m_1$ of times on the same network to obtain the averages, but it is also possible to average on a number $m_2$ of different realizations of the network itself. The total number, $m_T = m_1 \times m_2$, of samples thus obtained is used to calculate the probability $W_{M}(\theta)$, as it was shown previously, except that now the probability is associated to the network number of sites $M$ instead of certain characteristic length of the system.

Following the same scheme as before, we studied the RSA of dimers on ER and RR networks of $M$ sites (an $M$-lattice) considering different values of $M$ and different values of the degree $g$. The probability curves of $W_{M}$ are shown in Fig. \ref{wlynujerdos}(a) for values of $M$ between $10^3$ and $10^5$. Then, we look for $(dW'_{M}/d\theta)_{max}$ as a function of $M$ and $\Delta'_{M}$ as a function of $M$ [Eq. (\ref{nodos})]. Results are shown, in a log-log scale, in Figs. \ref{wlynujerdos}(b) and (c) and the fitting slopes collected in Table III. In each case, the obtained results support the dependence on the square root of $M$.

\begin{table}
\label{T2}
\begin{center}
\caption{Results collected from the linear fits of the ER and RR models.}
\begin{tabular}{p{1cm} |p{1cm} |p{3.6cm} |p{3.6cm} }

Graph & $\langle g \rangle$ & slope from $(dW'_{M}/d\theta)_{max}$ & slope from $\Delta'_{M}$   \\
\hline
ER	& 2 & 0.51(2) & -0.55(5)  \\
	& 4 & 0.50(5) & -0.53(3) \\
	& 5 & 0.48(5) & -0.54(6)  \\
\hline
RR	& 2 & 0.47(3) & -0.51(1)  \\
	& 4 & 0.47(3) & -0.52(2)  \\
	& 5 & 0.47(4) & -0.52(7)  \\
\end{tabular}
\end{center}
\end{table}

%
%
%
%
%

\begin{figure}
\begin{center}
\includegraphics[width=0.6\columnwidth]{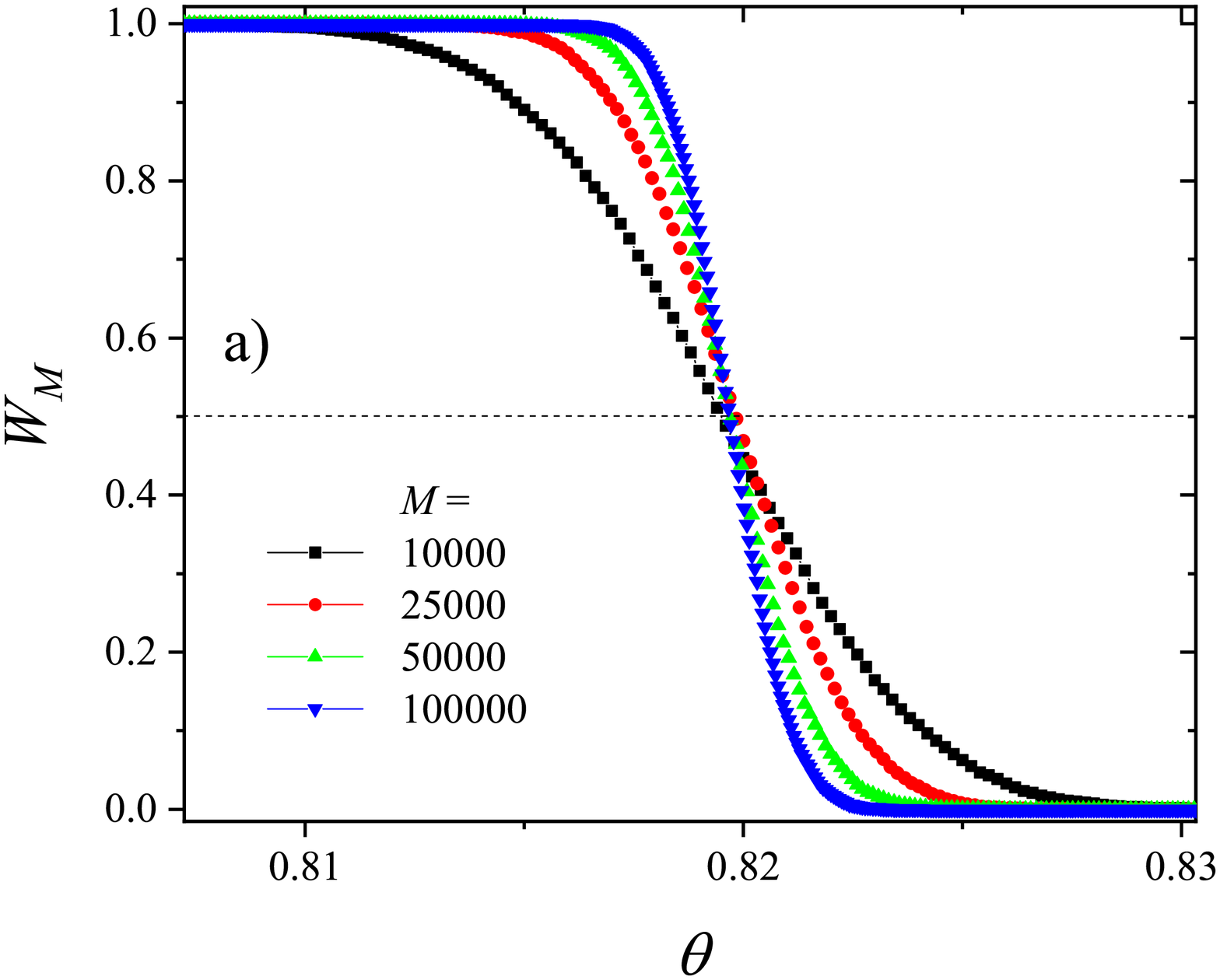}
\includegraphics[width=0.6\columnwidth]{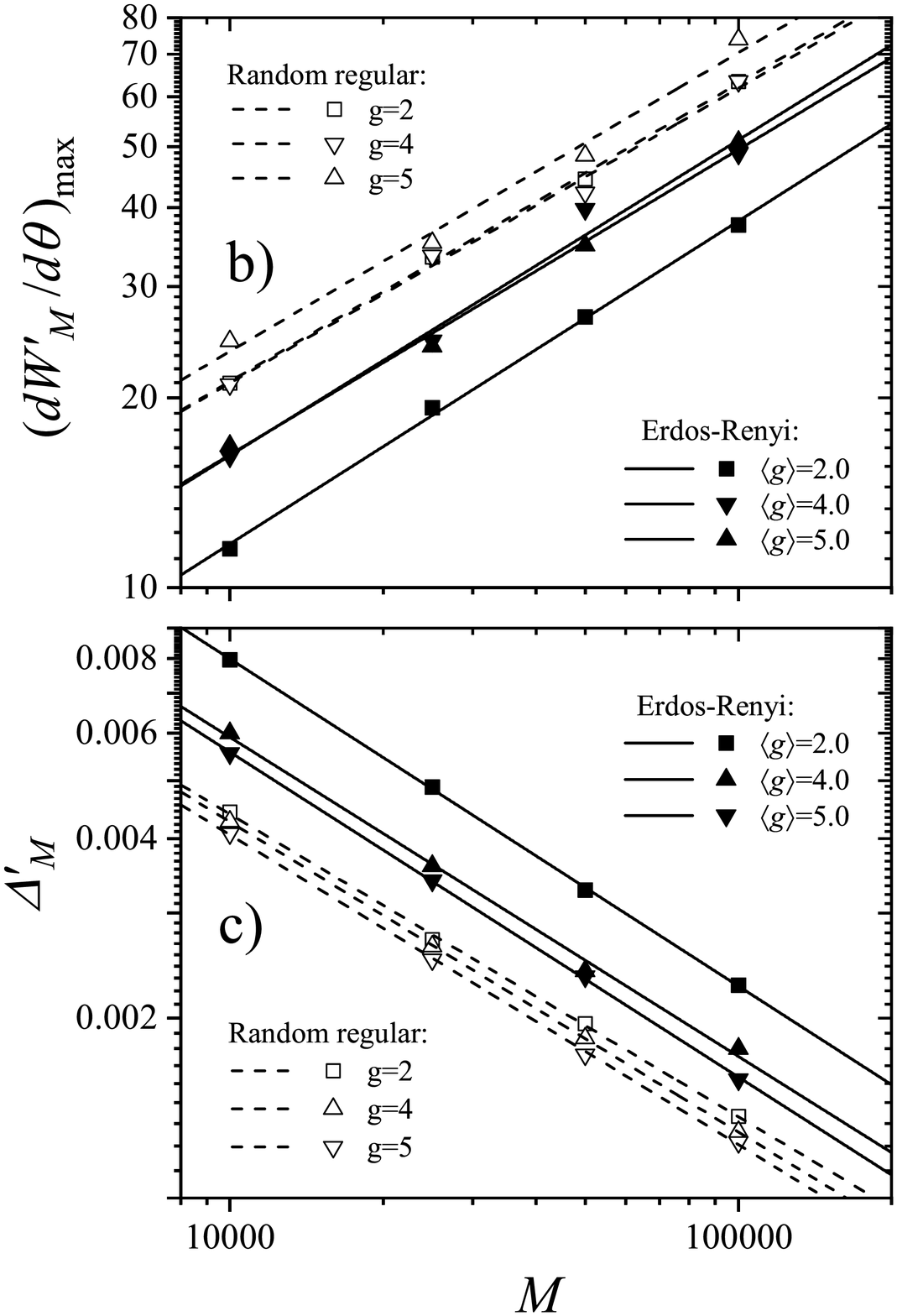}
\caption{(a) Curves of the jamming probability $W'_{M}$ as a function of the fraction of occupied sites $\theta$ for a typical Erdos-Renyi case with $M=10000, 25000, 50000, 100000$. Log-log plot of $(dW'_{M}/d\theta)_{\rm max}$ (a) and $\Delta'_{M}$ (b) as a function of $M$ for different cases, as indicated. \label{wlynujerdos}}
\end{center}
\end{figure}

Finally, Fig. \ref{todas} is presented as a summary of the systems addressed. This time all the data points have been represented as a function of the total number of sites, regardless of the type or size of the network considered.

\begin{figure}
\begin{center}
\includegraphics[width=0.8\columnwidth]{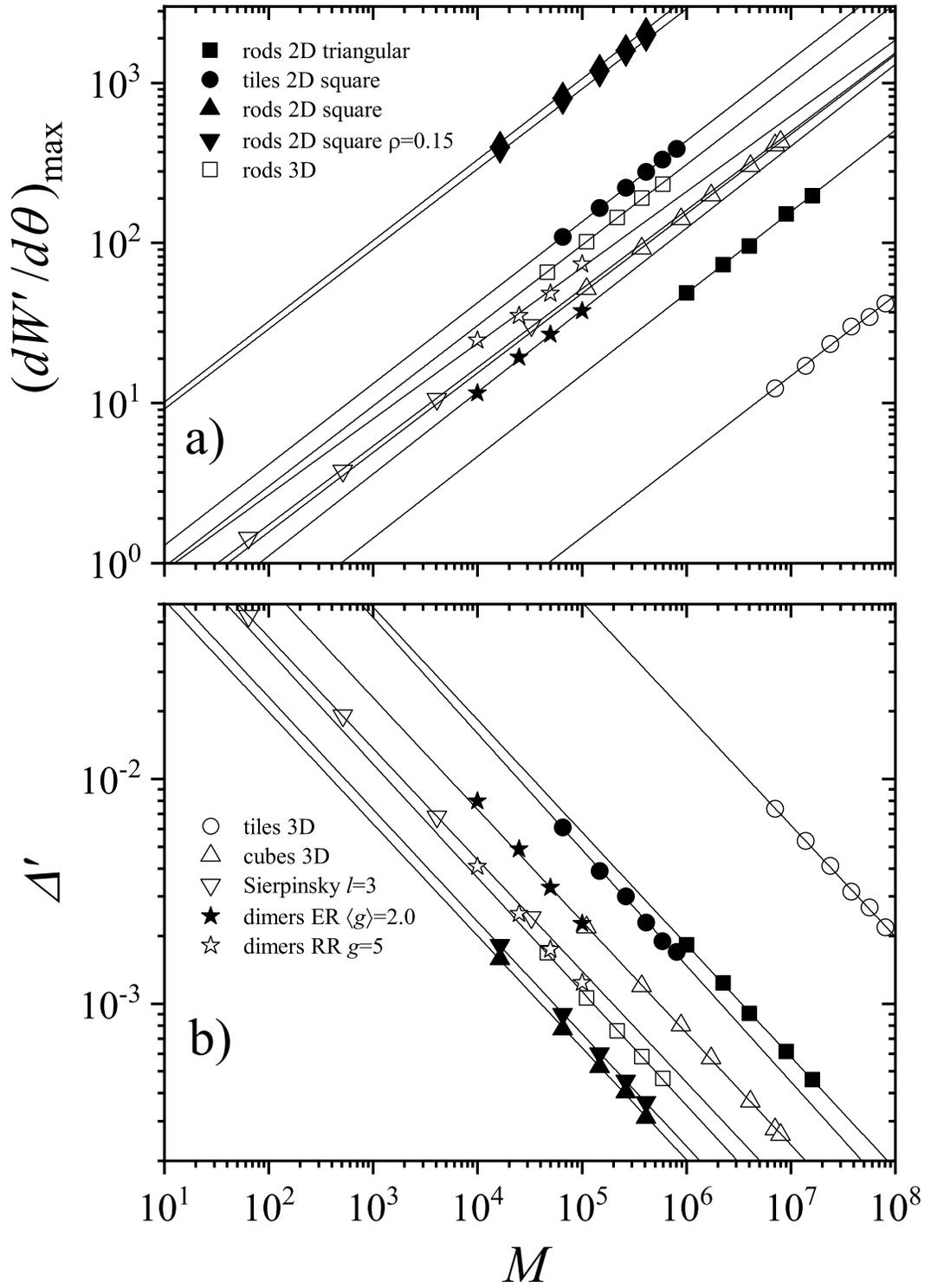}
\caption{(a) Log-log plot of $(dW'/d\theta)_{\rm max}$ as a function of $M$ for different RSA processes on Euclidean, fractal and random lattices, as indicated. (b) Same as part (a) for $\Delta'_{M}$. \label{todas}}
\end{center}
\end{figure}

\section{Conclusions}\label{conclu}

In this paper, irreversible adsorption of objects of different shapes and sizes on regular and random lattices has been studied. The adsorption dynamic was modeled by using the random sequential adsorption (RSA) algorithm. The process was monitored by following the behavior of the probability $W_{L(M)}(\theta)$ that a lattice composed of $L^d(M)$ elements reaches a coverage $\theta$.

By fitting $W_{L(M)}(\theta)$ $[\mathrm{d} W_{L(M)}(\theta)/\mathrm{d} \theta]$ with the error [Gaussian] function, the maximum of the derivative of the jamming probability $[\mathrm{d} W_{L(M)}(\theta)/\mathrm{d} \theta]_{max}$ and the width of the transition $\Delta_{L(M)}$ were obtained. These quantities are expected to behave asymptotically as $L^{1/\nu_j}$, where $\nu_j$ is the exponent characterizing the transition \cite{Vandewalle}.

In a first stage, the study was concentrated on the deposition of objects of different shapes and sizes on one-, two-, and three-dimensional Euclidean lattices. The results revealed that $\nu_j$ shows a simple dependence on the dimensionality of the lattice $d$: $\nu_j=2/d$. Thus, $\nu_j=2$, $1$ and $2/3$ for 1D, 2D and 3D lattices, respectively. The values of $\nu_j$ do not depend on size and shape of the depositing objects. Identical results were reported by Nakamura \cite{Nakamura} and by Vandewalle et al. \cite{Vandewalle} for the case of $k^2$-mers and rods on 2D square lattices.

The analysis was then extended to fractal lattices having dimension $d$ between 1 and 2. For this purpose, Sierpinski carpets were used as substrates and dimers as depositing objects. The tendency found for Euclidean lattices was confirmed. Thus, for Euclidean and fractal lattices, where the dimension $d$ and a characteristic length $L$ can be defined, the exponent $\nu_j$ results $\nu_j=2/d$. This property allowed us to rewrite the asymptotic functionality of $[\mathrm{d} W_{L(M)}(\theta)/\mathrm{d} \theta]_{max}$ and $\Delta_{L(M)}$ as $L^{1/\nu_j}=L^{d/2} = M^{1/2}$.

Finally, the study was generalized to RSA of dimers on the Erdos-Renyi  and Random Regular graphs. In this kind of systems, which are described solely in terms of number of sites and connectivity, the asymptotic law $M^{1/2}$ proved to be also valid.

\section{ACKNOWLEDGMENTS}

This work was supported in part by CONICET (Argentina) under project number PIP 112-201101-00615; Universidad Nacional de San Luis (Argentina) under project No. 03-0816; and the National Agency of Scientific and Technological Promotion (Argentina) under project  PICT-2013-1678. The numerical work were done using the BACO parallel cluster (http://cluster\_infap.unsl.edu.ar/wordpress/) located  at Instituto de F\'{\i}sica Aplicada, Universidad Nacional de San Luis - CONICET, San Luis, Argentina.

\end{large}

\end{document}